\RequirePackage{fix-cm}
\documentclass[twocolumn,epjc3]{svjour3}  
\smartqed  
\RequirePackage{graphicx}
\usepackage{siunitx}
\usepackage{xspace}

\newcommand{\micro}{\ensuremath{\mu}}

\newcommand{\bbonu}{\ensuremath{\beta\beta0\nu}}
\newcommand{\bbtnu}{\ensuremath{\beta\beta2\nu}}

\newcommand{\Qbb}{\ensuremath{Q_{\beta\beta}}}

\newcommand{\KR}{\ensuremath{^{83\textrm{m}}\mathrm{Kr}}\xspace}

\newcommand{\XE}{\ensuremath{{}^{136}\rm Xe}}

\newcommand{\TL}{\ensuremath{{}^{208}\rm{Tl}}}

\newcommand{\THO}{\ensuremath{{}^{228}{\rm Th}}}

\DeclareSIUnit\c{\mbox{$c$}}
\DeclareSIUnit\magn{\mbox{$\times$}}
\DeclareSIUnit\min{min}
\DeclareSIUnit\week{week}
\DeclareSIUnit\year{yr}
\DeclareSIUnit\years{years}
\DeclareSIUnit\yr{yr}
\DeclareSIUnit\standard{std}
\DeclareSIUnit\str{sr}
\DeclareSIUnit\ppm{ppm}
\DeclareSIUnit\ppb{ppb}
\DeclareSIUnit\ppt{ppt}
\DeclareSIUnit\pe{PE}
\DeclareSIUnit\spe{SPE}
\DeclareSIUnit\ev{events}
\DeclareSIUnit\ct{counts}
\DeclareSIUnit\neutron{\mbox{$n$}}
\DeclareSIUnit\smp{samples}
\DeclareSIUnit\Sample{S}
\DeclareSIUnit\ch{ch}
\DeclareSIUnit\hit{hit}
\DeclareSIUnit\hits{hits}
\DeclareSIUnit\bin{(\mbox{5-PE}~bin)}
\DeclareSIUnit\sgm{\mbox{$\sigma$}}
\DeclareSIUnit\rms{RMS}
\DeclareSIUnit\keVr{\mbox{keV$_{\rm nr}$}}
\DeclareSIUnit\keVee{\mbox{keV$_{e{\rm e}}$}}
\DeclareSIUnit\ph{photon}
\DeclareSIUnit\pes{pes}
\DeclareSIUnit\el{electrons}
\DeclareSIUnit\pm{PMT}
\DeclareSIUnit\inch{"}
\DeclareSIUnit\bit{bit}
\DeclareSIUnit\sample{samples}
\DeclareSIUnit\barn{barn}
\DeclareSIUnit\bara{bar}
\DeclareSIUnit\barg{barg}
\DeclareSIUnit\mlardepth{\mbox(meter~of~\LAr~depth)}
\DeclareSIUnit\Curie{Ci}
\DeclareSIUnit\psi{psi}
\DeclareSIUnit\parsec{pc}
\DeclareSIUnit\liveday{\mbox{live-days}}
\DeclareSIUnit\days{\mbox{days}}
\DeclareSIUnit\day{\mbox{day}}
\DeclareSIUnit\miles{\mbox{miles}}
\DeclareSIUnit\degreeC{\mbox{$^{\circ}$C}}
\DeclareSIUnit\electron{\mbox{$e^-$}}
\DeclareSIUnit\Euro{\mbox{\euro}}
\DeclareSIUnit\cph{cph}
\DeclareSIUnit\neq{neq}
\DeclareSIUnit\unit{unit}
\DeclareSIUnit\byte{Byte}
\DeclareSIUnit\Bq{\becquerel}

\usepackage{amsmath}
\usepackage[colorlinks=true, linkcolor=blue,citecolor=blue]{hyperref}
\journalname{Eur. Phys. J. C}
\begin{document}

\title{Demonstration of Sub-Percent Energy Resolution in the NEXT-100 Detector
}

\author{%
M.~P\'erez Maneiro\thanksref{addr22}
\and
M.~Mart\'inez-Vara\thanksref{addr16}
\and
S.~Torelli\thanksref{e1,addr19}
\and
G.~Mart\'inez-Lema\thanksref{e2,addr19}
\and
P.~Novella\thanksref{addr16}
\and
J.A.~Hernando~Morata\thanksref{addr22}
\and
J.J.~G\'omez-Cadenas\thanksref{e3,addr19,addr8}
\and
C.~Adams\thanksref{e4,addr2}
\and
H.~Almaz\'an\thanksref{addr19}
\and
V.~\'Alvarez\thanksref{addr23}
% ...
}

\thankstext{e1}{Corresponding author}
\thankstext{e2}{Now at Instituto de Física Corpuscular, Spain}
\thankstext{e3}{NEXT Spokesperson}
\thankstext{e4}{Now at NVIDIA}
\thankstext{e5}{Deceased}
\thankstext{e6}{Now at Weizmann Institute of Science, Israel}
\thankstext{e7}{On leave}
\thankstext{e8}{Now at University of North Carolina, USA}
\thankstext{e9}{Now at Los Alamos National Laboratory, USA}

\institute{
% ...
Argonne National Laboratory, Argonne, IL 60439, USA \label{addr2}
% ...
\and
Ikerbasque (Basque Foundation for Science), Bilbao E-48009, Spain \label{addr8}
% ...
\and
Instituto de F\'isica Corpuscular (IFIC), CSIC \& Universitat de Val\`encia, Paterna E-46980, Spain \label{addr16}
% ...
\and
Donostia International Physics Center (DIPC), San Sebasti\'an E-20018, Spain \label{addr19}
% ...
\and
Instituto Gallego de F\'isica de Altas Energ\'ias, Universidade de Santiago de Compostela, Santiago de Compostela E-15782, Spain \label{addr22}
\and
Instituto de Instrumentaci\'on para Imagen Molecular (I3M), CSIC–UPV, Valencia E-46022, Spain \label{addr23}
}

\date{Received: date / Accepted: date}

\maketitle
\setcounter{tocdepth}{2}
\tableofcontents
\vspace{.7cm}

\begin{abstract}
NEXT-100 is a high-pressure xenon time projection chamber with electroluminescent amplification, designed to operate with up to $\sim$70.5 kg at 13.5 bar. It is the most recent detector developed by the NEXT collaboration to search for the neutrinoless double-beta decay (\bbonu) of \XE. The NEXT gas TPC technology offers the best energy resolution near the Q-value of the decay (\Qbb\ = 2458 keV) among xenon detectors, which is set by design to be $<$1\% FWHM. We report here the high-energy calibration of the detector using a \THO\ source, demonstrating excellent linear response and an energy resolution of $(0.90 \pm 0.02)\%$ FWHM at the $^{208}$Tl photopeak (2615 keV). This performance extrapolates to a resolution at the double-beta decay end-point of $R(Q_{\beta\beta}) = (0.93 \pm 0.02)\%$ FWHM, confirming the detector’s capability for precision energy measurement in the search for \bbonu.
\end{abstract}

\section{Introduction}\label{s:intro}
Neutrinoless double beta decay (\bbonu) is a hypothesized lepton-number-violating process in which two neutrons in a nucleus simultaneously decay into two protons, emitting two electrons. Its experimental signature is a monoenergetic peak at the decay $Q$-value, with events featuring two electron tracks exhibiting Bragg peaks at both ends. For the search of this rare decay, the NEXT experiment has developed a new High-Pressure Electroluminescent xenon TPC.

One of the key features of the NEXT technology is its excellent intrinsic energy resolution of 0.3\% FWHM at the $^{136}$Xe $Q$-value (2458 keV), a crucial ingredient for neutrinoless double beta decay (\bbonu) searches. Demonstration of such good energy resolution was achieved with the prototypes NEXT-DEMO and NEXT-DBDM \cite{NEXT:2012lrw,2013JInst} and confirmed by the NEXT-White demonstrator, which operated at the Canfranc Underground Laboratory (LSC) from 2016 to 2021 \cite{NEXT:2018rgj} and measured an energy resolution of $R = (0.91 \pm 0.07)\%$ at the Q-value of the decay \cite{Renner_2018,Resowhite}. Exploiting this energy resolution and the topological discrimination of the tracks~\cite{DackRejNN,EvIDNEXTW,NEXT:2021vzd}, NEXT-White also characterized the backgrounds associated to the NEXT technology~\cite{BkgNW1,BkgNW2}, and performed the first double beta decay searches~\cite{PhysRevC.105.055501,NEXT:2023daz}.

NEXT-White was superseded by the NEXT-100 detector, presently taking data at the LSC. NEXT-100 is designed to hold up to $\sim$ 70 kg of xenon at a nominal pressure of 13.5 bar. Initial operation of the chamber is currently conducted at 4 bar, while operation at nominal pressure is foreseen to start in early 2026. The commissioning of the apparatus has shown excellent performance and stability, including, in particular, an electron lifetime on the order of tens of ms, exceeding the maximum drift time by more than an order of magnitude~\cite{NEXT:2025yqw}. Subsequently, a low-energy calibration campaign using $^{83\textrm{m}}$Kr demonstrated excellent energy resolution, achieving $R = 4.37\ \%$ FWHM at 41.5 keV and 4 bar, and demonstrated the effectiveness of the geometrical light-response corrections~\cite{NEXT:2025fpq}.

In this work, we present the results of the high-energy calibration of the detector using a $^{228}$Th radioactive source. This source produces multiple monochromatic gamma lines spanning a broad energy range, among which the 2615~keV gammas from $^{208}$Tl, present in the $^{228}$Th decay chain. The corresponding gamma peaks have been used to characterize the detector response and evaluate its energy resolution. Particular emphasis is placed on the $^{208}$Tl photopeak, which lies very close to the $Q_{\beta\beta}$ value and therefore provides crucial information for extrapolating the expected resolution at the energy relevant for neutrinoless double-beta decay searches.

The manuscript is organized as follows. We start with an overview of the detector and its operating conditions during the calibration runs (Section ~\ref{s:n100}). In Section \ref{s:he_calib} we describe the data analysis, detailing the event selection, the geometrical and lifetime corrections, as well as energy-scale calibration. Section~\ref{s:ene_r} presents the energy resolution achieved by analyzing the peaks produced by the gamma rays from the calibration source, and discusses the extrapolation of the resolution to the \Qbb\ energy. In Section~\ref{sec.photopeak}, we apply an additional single-track selection to evaluate detector performance under conditions relevant for neutrinoless and two-neutrino double-beta decay (\bbtnu) analyses. Section~\ref{sec:sep} demonstrates the power of the topological discrimination capability, which enables the identification of the single-escape peak, otherwise buried in the background. The results of this study are summarized, discussed and concluded in Section \ref{s:summ}.

\section{The NEXT-100 detector}\label{s:n100}
\subsection{Detector description}\label{ss:det}

\begin{figure*}[t]
    \centering
    \includegraphics[width=.6\textwidth]{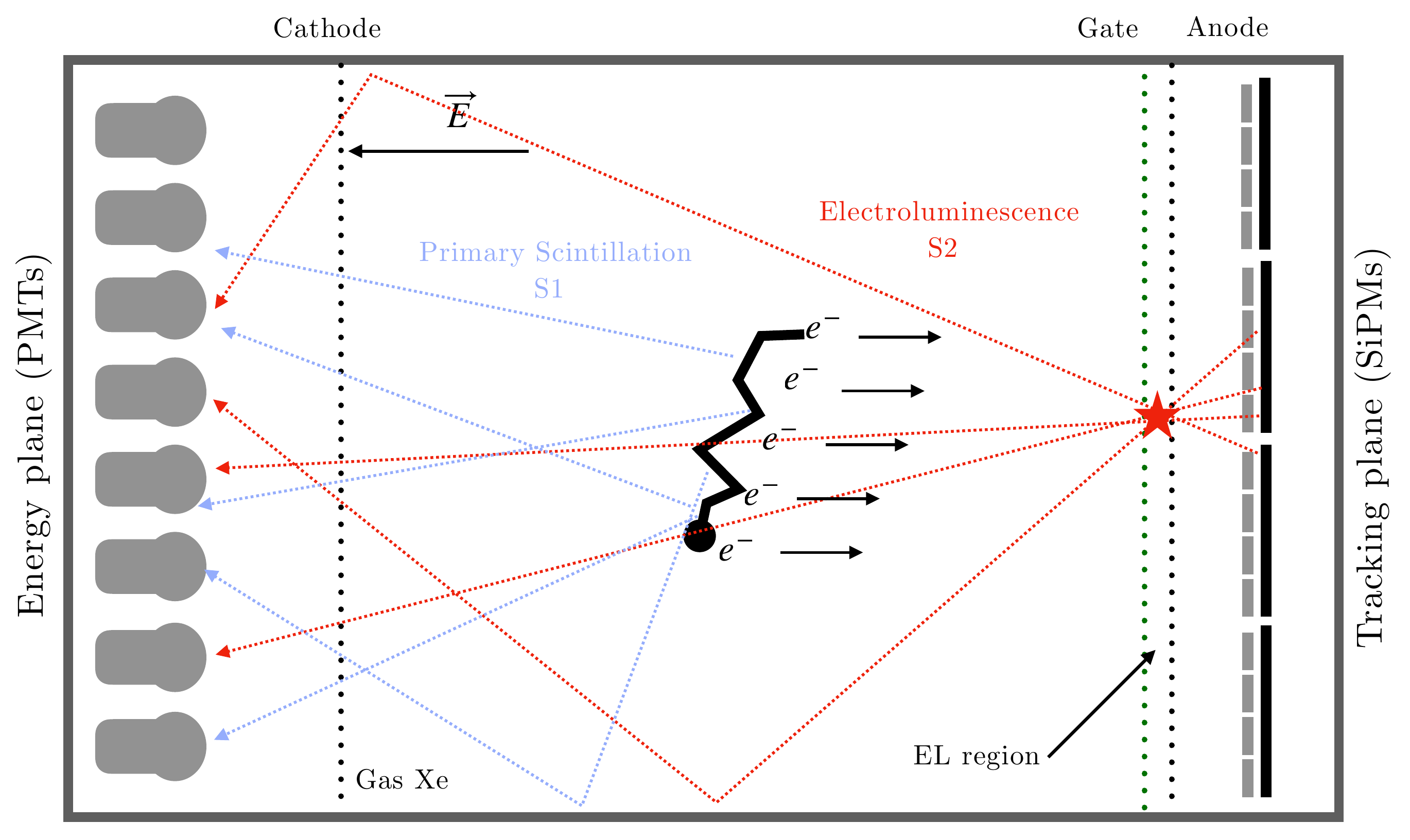}
    \caption{Principle of operation of the NEXT-100 detector.}
    \label{fig:principle}
\end{figure*}

The NEXT-100 apparatus~\cite{NEXT:2025yqw} is the latest and largest experiment of the NEXT detector series. Like its predecessors, it is an asymmetric high-pressure xenon TPC, whose operating principle is illustrated in Figure~\ref{fig:principle}. When an charged particle, such as a high-energy electron, interacts in the chamber, it generates a trail of ionization
electrons along its path, together with excited xenon atoms. Xenon, de-exciting, emits vacuum ultraviolet (VUV) scintillation light with a peak wavelength of 172~nm. This prompt scintillation burst, referred to as S1, lasts a few hundred nanoseconds and is recorded by a plane of photomultiplier tubes (PMTs)—the energy plane (EP)—located behind a transparent cathode grid held at negative high voltage. Detection of S1 defines the event start time, $t_0$.

A moderate electric field (a few hundred V/cm) drifts the ionization electrons toward the anode region, which is formed by two conductive transparent grids separated by a small gap, known as the electroluminescence (EL) region. A strong high voltage is applied across this gap, creating the intense electric field required to generate electroluminescence light. In this region, the electrons are accelerated enough to excite xenon atoms between collisions, producing a proportional signal of secondary scintillation light (S2). This nearly fluctuationless amplification, enables excellent energy resolution by converting essentially all ionization charge into measurable light. 

The S2 light is detected both by the EP and by a plane of silicon photomultipliers (SiPMs) forming the Tracking Plane (TP), located 15 mm behind the anode. To enhance photon collection efficiency, the TPC’s inner surface is lined with PTFE panels coated with a $\sim$3-4~$\micro$m layer of tetraphenyl butadiene (TPB), a wavelength shifter that converts VUV photons into the blue region, where they are more efficiently detected by the PMTs and SiPMs.

In NEXT-100, the TPC has a drift length of 1187 mm, a diameter of 983 mm, and an EL gap of 9.70 mm. The EP consists of 60 Hamamatsu R11410-10 PMTs, of which 48 were fully operational during the data taking. Each PMT is enclosed in an individual, vacuum-insulated capsule and optically coupled to the active volume through a sapphire window. The PMTs are arranged in a concentric hexagonal (honeycomb) pattern. The TP comprises 3584 Hamamatsu S13372-1350TE SiPMs ($1.3\times1.3$~mm$^2$ each), arranged in modular boards of $8\times8$ sensors with a 15.55~mm pitch. The SiPM array extends beyond the active region to minimize edge losses. Both the sapphire windows coupling the PMTs and the SiPM surfaces are coated with TPB.

Signals from the EP and TP provide complementary measurements of S2. The combination of spatial information from the SiPM array (transverse coordinates) and the drift time (longitudinal coordinate) enables full 3D reconstruction of particle tracks. The SiPMs are sampled at DAQ level with a sampling width of 1~$\micro$s (corresponding to $\sim$1~mm along the drift direction), while the faster PMT waveforms, originally sampled at 25 ns, are rebinned to match the SiPM sampling rate. 

\begin{figure*}[t]
    \centering
    \includegraphics[
        width=.7\textwidth
    ]{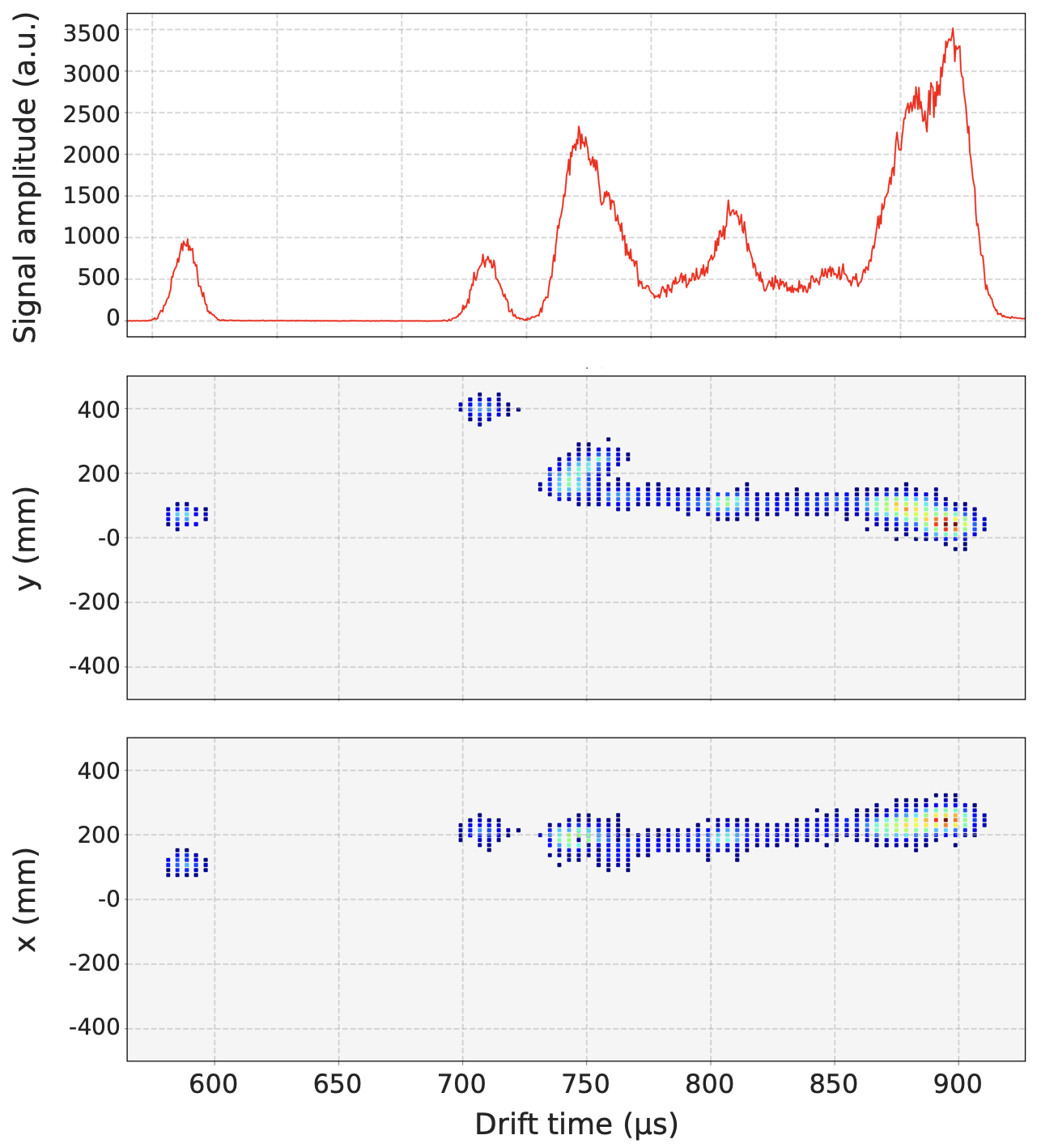}
    \caption{Example of PMT waveforms associated to tracks in an event. Top: Sum of the PMT waveform for an event. The time axis corresponds to the raw time recorded by the DAQ system within its 2 ms acquisition window. Center/Bottom: y–z and x–z projections reconstructed from SiPM data; the time axis represents the drift time measured from the S1 signal.}
    \label{fig:a_track}
\end{figure*}

Figure~\ref{fig:a_track} shows an example PMT waveform together with the corresponding 2D projections of the reconstructed SiPM hits, representing the tracks. The different clusters of light corresponding to the different tracks can be seen. The correlation between the PMT waveform and the localized energy deposition in the SiPM clusters can be seen, with the waveform peak aligned with the Bragg peak of the track.

\subsection{Operational conditions}\label{ss:runs}
\begin{table}[tbhp!]
\centering
\begin{tabular}{|ll|}
\hline
\multicolumn{1}{|l|}{Detector parameter} & Value                    \\ \hline
Pressure                                 & $\sim$4 bar             \\
V$_{\mathrm{cathode}}$                            & -23.0 kV                 \\
V$_{\mathrm{gate}}$                               & -8.8 kV                  \\
EL reduced field                         & 2.2 kV/cm/bar            \\
Drift field                              & 120 V/cm \\
Xe active mass                           & $\sim$19 kg                    \\ \hline
\end{tabular}
\caption{Operating conditions of the detector during the high-energy calibration run.}
\label{tab:opconditions}
\end{table}
The data used in this analysis were collected during the first half of July 2025 and consist of two $\sim$24~h runs with \KR\ and eight runs with a \THO\ source, for a total live time of approximately 182~h. The detector operating conditions for these runs are summarized in Table~\ref{tab:opconditions}. 

The \KR\ source is produced from small zeolite beads containing $^{83}$Rb integrated into the gas system. The decay of $^{83}$Rb generates \KR, which mixes with the xenon and circulates uniformly throughout the active volume. The subsequent decay of \KR\ yields point-like energy depositions of 41.5~keV. As detailed in section~\ref{s:he_calib}, these events are used to build the detector response map and to correct for electron lifetime variations and geometrical effects \cite{Mart_nez_Lema_2018}.

The \THO\ data were acquired with the source located in a dedicated calibration port, which is positioned outside the pressure vessel but inside the NEXT-100 lead shielding. Specifically, the port is positioned at (395 mm, 395 mm, 534 mm) in the xyz coordinates. The \THO\ source produces several characteristic features in the energy spectrum: the 511~keV line from positron–electron annihilation very close to a gamma line at 510.7 keV from $^{208}$Tl, the 583 keV, 727 keV, and 860~keV $\gamma$ lines from $^{208}$Tl and $^{212}$Bi, the 2615~keV photopeak from \TL\ $\gamma$ rays, and the corresponding single- and double-escape peaks (SEP and DEP) at 2104 keV and 1593 keV, respectively. These spectral features are used to calibrate and characterize the detector response, evaluate the energy resolution, and extrapolate its expected performance at \Qbb. 

The detector’s dual-trigger acquisition system enables the simultaneous collection of low-energy and high-energy data through parallel channels with independent trigger configurations. This allows \KR\ data to be recorded concurrently with \THO\ runs, providing a stable reference for monitoring variation in light yield over time intervals of a few hours. 
\section{Detector response calibration}\label{s:he_calib}
\subsection{Event selection}\label{ss:ev_selection}
The event selection requires a well-defined S1 signal, followed in time by a single S2 pulse. Subsequently, a set of geometrical cuts is applied to restrict reconstructed tracks to the fiducial volume and ensure full containment. These include longitudinal (z) cuts to exclude events near the cathode or the EL region, as well as a maximum radial cut to remove tracks approaching the detector walls, where border effects may degrade the energy resolution.

\begin{figure*}[t]
    \centering
    \includegraphics[width=0.95\textwidth]{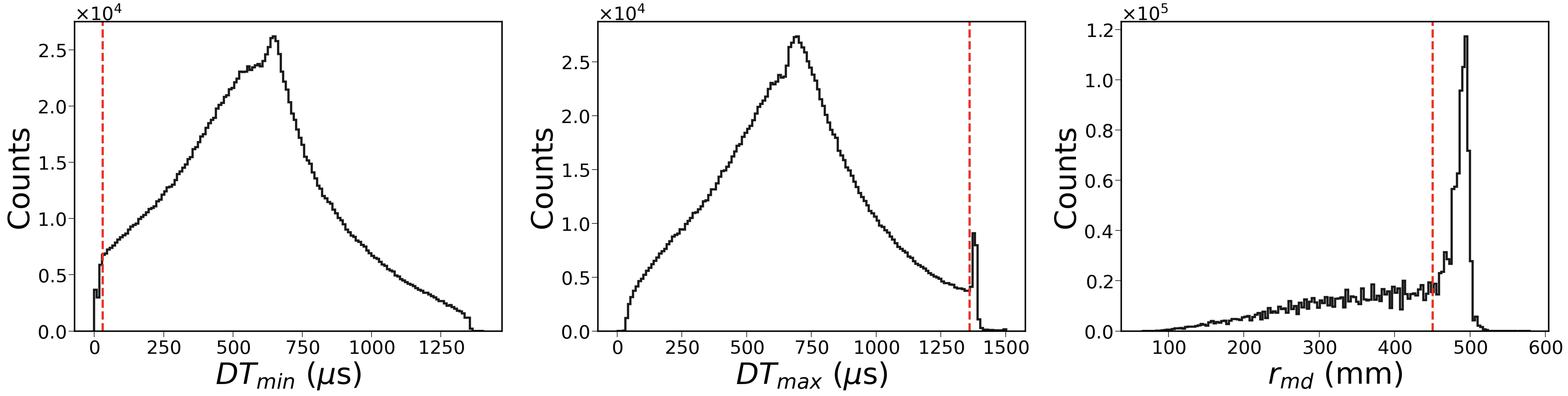}
    \caption{Left/center: Distributions of the minimum and maximum value of the track drift times within the events. Right: distribution of the maximum radial distance of the tracks. The red lines represent the selection cut in the three distributions.}
    \label{fig:geo_distr}
\end{figure*}

Specifically, the selection criteria applied are:
\begin{itemize}
    \item The number of S1 and S2 signals in the waveform is required to be $N_{S1}=N_{S2}=1$. 
    \item $\mathrm{DT_{\min}} > 20~\micro\mathrm{s}$ and $DT_{\max} < 1350~\micro\mathrm{s}$, where $DT_{\min}$ and $DT_{\max}$ represent the minimum and maximum drift times (DT) value of the tracks within an event.
    \item $r_{md} < 450$~mm, where $r_{md}$ is the maximum radial distance of the tracks in an event from the center of the xy plane.
\end{itemize}
These cuts define the detector fiducial volume. The validity of these selection cuts is illustrated in Figure \ref{fig:geo_distr}, which shows the distributions of $DT_{\min}$ (left panel), $DT_{\max}$ (central panel), and $r_{md}$ (right panel). The applied cuts are indicated by red lines. The first two distributions peak near the detector center, consistent with the source being positioned along the central region of the z-axis. A sharp feature appears near the cathode, corresponding to tracks crossing it (not fully contained). The radial distribution shows a significant rise in event density near the edge, primarily due to low-energy gammas interacting close to the source.

\subsection{Geometric and Temporal Light Variation Effects}
The collection of scintillation light by the PMTs varies with the $(x,y)$ position of the emission along the electroluminescence plane. This dependence arises from changes in the solid angle subtended by each PMT at different locations, as well as from reflections on the detector barrel. In addition, electrons drifting through the gas have a finite probability of being captured by impurities before reaching the anode. Consequently, a track produced at a given $z$ position yields a reduced charge at the anode. 
The combined effect of these spatial and lifetime inhomogeneities leads to a position-dependent light response throughout the detector volume.
To correct for these spatial and lifetime effects, data from the \KR\ runs are used to construct a three-dimensional map of the local detector response as a function of position $(x, y, z)$. The active volume is divided into $100\times100\times10$ bins, and the average \KR\ light output is computed in each bin. The resulting map is then normalized to the mean value of the first $z$ slice within a radius of 100~mm, yielding the correction coefficients $C_{k,l,m}$, with $k,l,m$ being the index of the map bins. Applying this position-dependent correction to \KR\ data homogenizes the detector light response and improves the overall energy resolution \cite{Mart_nez_Lema_2018}.

For extended tracks, this correction is applied using the local charge information provided by the SiPM tracking plane. The SiPM position combined with the drift time information give us a set of $(x,y,z)$ points, each associated with a measured charge. The PMT waveform is divided into 4~$\micro$s time slices, each containing a fraction $E^{wf}_{i}$ of the total event energy $E^{wf}$. The 4~$\micro$s time interval was chosen based on the SiPM sampling time of 1~$\micro$s, applying a rebinning factor of 4 to enhance the signal-to-noise ratio during track reconstruction. Considering $Q_{j,i}$ as the charge recorded in the voxel $j$ in correspondence to the slice $i$, an energy $E^{V}_{ji}$
\[
E^{V}_{ji} = E^{wf}_{i} \cdot \frac{Q_{ji}}{\sum_k Q_{ki}}
\]
is assigned to each voxel. For every voxel falling within the bin $(k,l,m)$ of the response map, the corresponding energy is divided by the correction factor $C_{k,l,m}$. The total corrected energy of the track is then obtained by summing over all voxels. For reference, see Figure~\ref{fig:a_track}.

In addition to spatial corrections, temporal variations in light yield are continuously monitored and corrected; these variations remain below 0.3\% over more than a week of operation, demonstrating excellent system stability. Corrections are applied using Kr data acquired alongside the high-energy events via the dual-trigger system. To equalize high-energy data acquired at different times, the light measured at time $t$ is rescaled by dividing by a normalization factor $N(t)$, defined as:
\[
N(t) = \frac{LO_{kr}(t)}{LO_{kr}(t_0)}
\]
where $LO_{Kr}(t)$ represents the light output measured with krypton at a time $t$, and $LO_{Kr}(t_0)$ is the corresponding light output measured with krypton at the reference time $t_0$.
This normalization corrects for temporal variations in detector response, ensuring that data taken at different times are brought to a common reference scale.

\subsection{Calibration of the energy scale}

\begin{figure*}[t]
    \centering
    \includegraphics[width=0.9\textwidth]{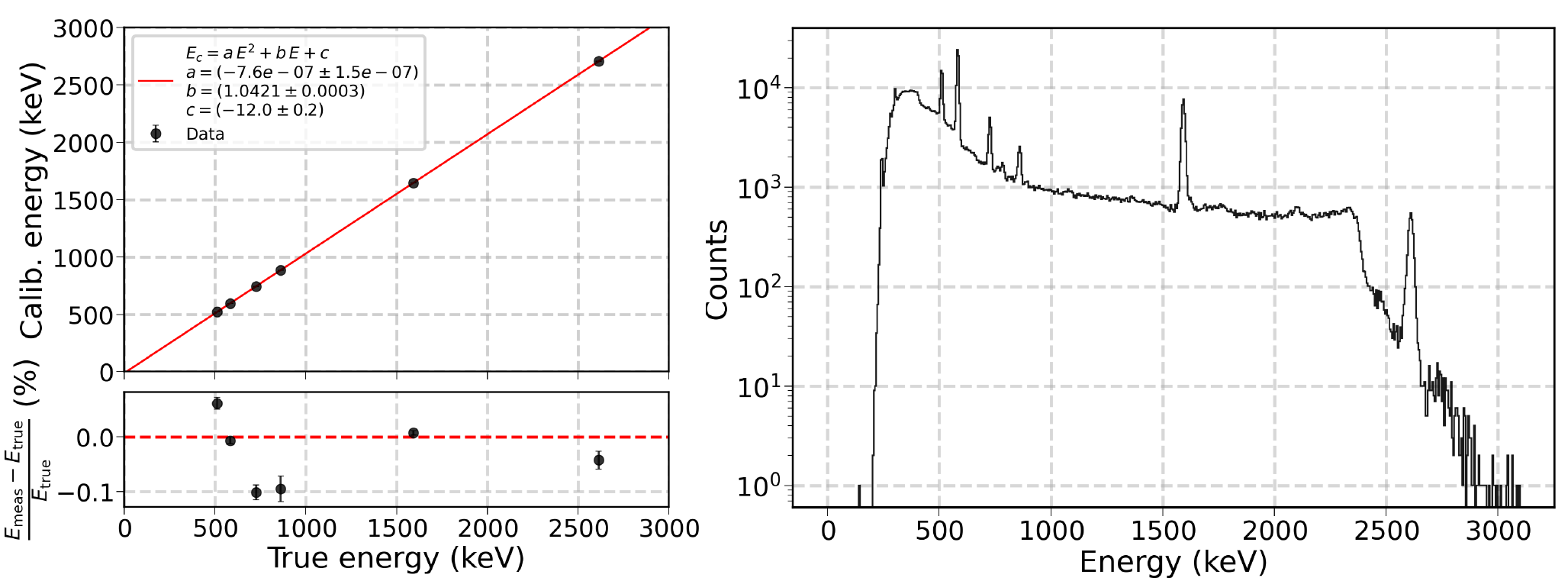}
    \caption{Left: Kr-calibrated energy as a function of the true energy. The data are fitted with a second-degree polynomial to account for the slight nonlinearity in the detector response. In the bottom panel, the relative energy deviation, expressed as a percentage, between the energy measured after this additional calibration and the true energy is reported. Right: Energy spectrum corrected for residual non-linearity, with all the peaks produced by the gammas emitted in the \THO\ chain visible.}
    \label{fig:spectrum}
\end{figure*}

After applying corrections for geometrical effects and temporal variations in light yield, the data were calibrated using the \KR\ ``standard candle'' to convert photoelectrons into (calibrated) energy $E_c$. The mean values of the identified peaks of the calibrated energy spectrum were then plotted as a function of their nominal energies ($E_{true}$) to evaluate the linearity of the detector response. The calibrated data were fitted with a second-degree polynomial, $E_c = a\cdot E_{true}^2 +b\cdot E_{true} +c$, to account for residual nonlinearities across the full energy range. The best-fit parameters are $a = (-7.9 \pm 1.4 )\times 10^{-7}$ keV$^{-1}$, $b = (1.0421 \pm 0.0003)$, and $c = (-12.0 \pm 0.1)$ keV. The fit and corresponding data points, together with the plot showing the deviation of the measured energy after this additional calibration and the true energy, are shown in the left panel of Figure~\ref{fig:spectrum}. The very small value of $a$ confirms the excellent linearity of the detector response. In this final calibration, the deviation between the measured and true energies remains below 0.1\% in the whole energy range.

After correcting for the residual non-linearity, the resulting energies—henceforth referred to simply as `Energy'—are summed over all tracks within each event to obtain the energy spectrum, which is shown in the right panel of Figure~\ref{fig:spectrum}.
\section{Energy resolution}\label{s:ene_r}

In an ideal detector, photons of the same energy would produce a delta-function–like peak in the energy spectrum, reflecting a perfectly monochromatic response. In real detectors, however, various effects such as statistical fluctuations in light production and collection, charge recombination, and electronic noise broaden this peak. The energy resolution $R$ of a peak quantifies this broadening and is defined as the ratio between the full width at half maximum (FWHM) of the peak and its mean energy.

\begin{figure*}[t]
    \centering
    \includegraphics[width=0.9\textwidth]{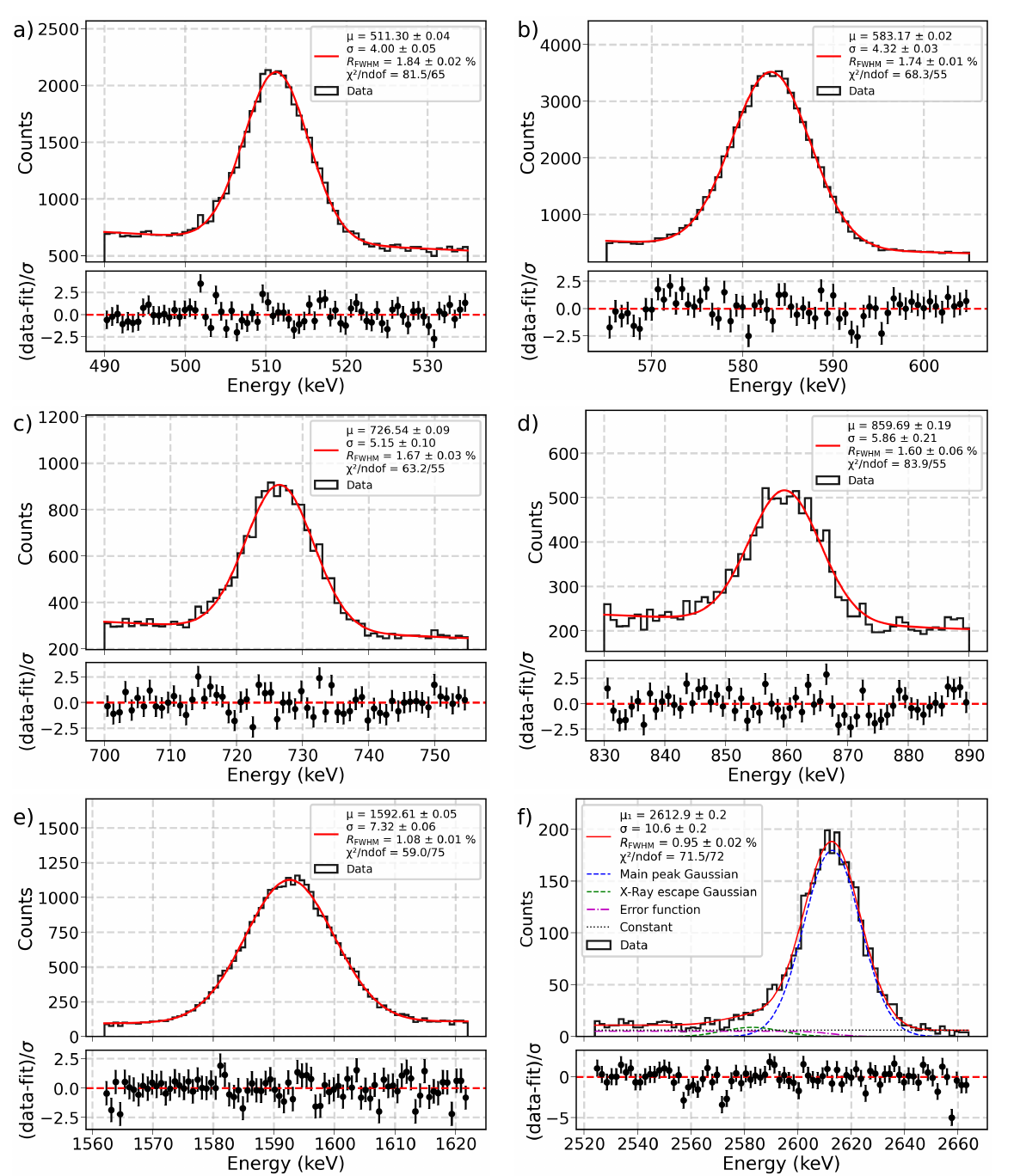}
    \caption{Energy distributions for the a) 511 keV, b) 583 keV, c) 727 keV, d) 860 keV, e) 1593 keV, and f) 2615~keV peaks, along with their corresponding fits. The legend reports the fit parameters and $\chi^2/\mathrm{ndof}$ values. Below each distribution, the residuals normalized to the standard deviation are displayed. For f) the legend explicitly displays all component functions used in the fit.}
    \label{fig:fits_le}
\end{figure*}

To evaluate the energy resolution, we select first events that satisfy the $N_{S1} = N_{S2} = 1$ condition on the waveforms. Following the calibration procedures described above, the energy spectra corresponding to the peaks listed earlier were fitted using appropriate analytical models. 
Figure~\ref{fig:fits_le} shows the fits to the 511 keV (a), 583 keV (b), 727 keV (c), 860 keV (d), 1593 keV (e), and 2615~keV (f) peaks, along with the corresponding fit parameters, residuals, and $\chi^2/\mathrm{ndof}$ values.

For most peaks, the fitting function consisted of an exponential term describing the Compton continuum and a Gaussian function modeling the peak itself. In the case of the DEP, a first-degree polynomial was used to account for the non-exponential shape of the local background. For the \TL\ photopeak, a more complex model was adopted, following Refs.~\cite{PhysRevLett.120.132501,Agostini_2021}:
\begin{equation}
\label{eq:fitppsc}
\begin{split}
f(x) ={} A\,e^{-\frac{(x-\mu_1)^2}{2\sigma^2}}
        &+ B\,e^{-\frac{(x-(\mu_1-30\ keV))^2}{2\sigma^2}} \\
       &+ C\,\mathrm{erfc}\!\left(\frac{-(x-\mu_1)}{\sqrt{2}\,\sigma}\right)
        + D .
\end{split}
\end{equation}
Here, the second Gaussian term (with fixed mean value) accounts for photoelectric events where a K-shell xenon X-ray is produced and escapes the active volume. This second Gaussian is necessary since the X-ray escape peak is immersed in the main peak. The situation is different for the photopeaks at lower energy, where the X-ray escape peak is fully separated from the main photopeak, and, as a matter of fact, masked in the continuous background. For example, the 511 keV peak has a resolution of 9.5 keV, and the 860 keV peak has a resolution of 14 keV, so both peaks are fully separated from the X-ray, 30 keV to the left. Finally, the complementary error function (sharing the same mean and width as the main Gaussian) models the low-energy edge produced by multiple Compton interactions, while the constant term represents the residual flat background beyond the peak.



\begin{figure}[t]
    \centering
    \includegraphics[width=.9\columnwidth]{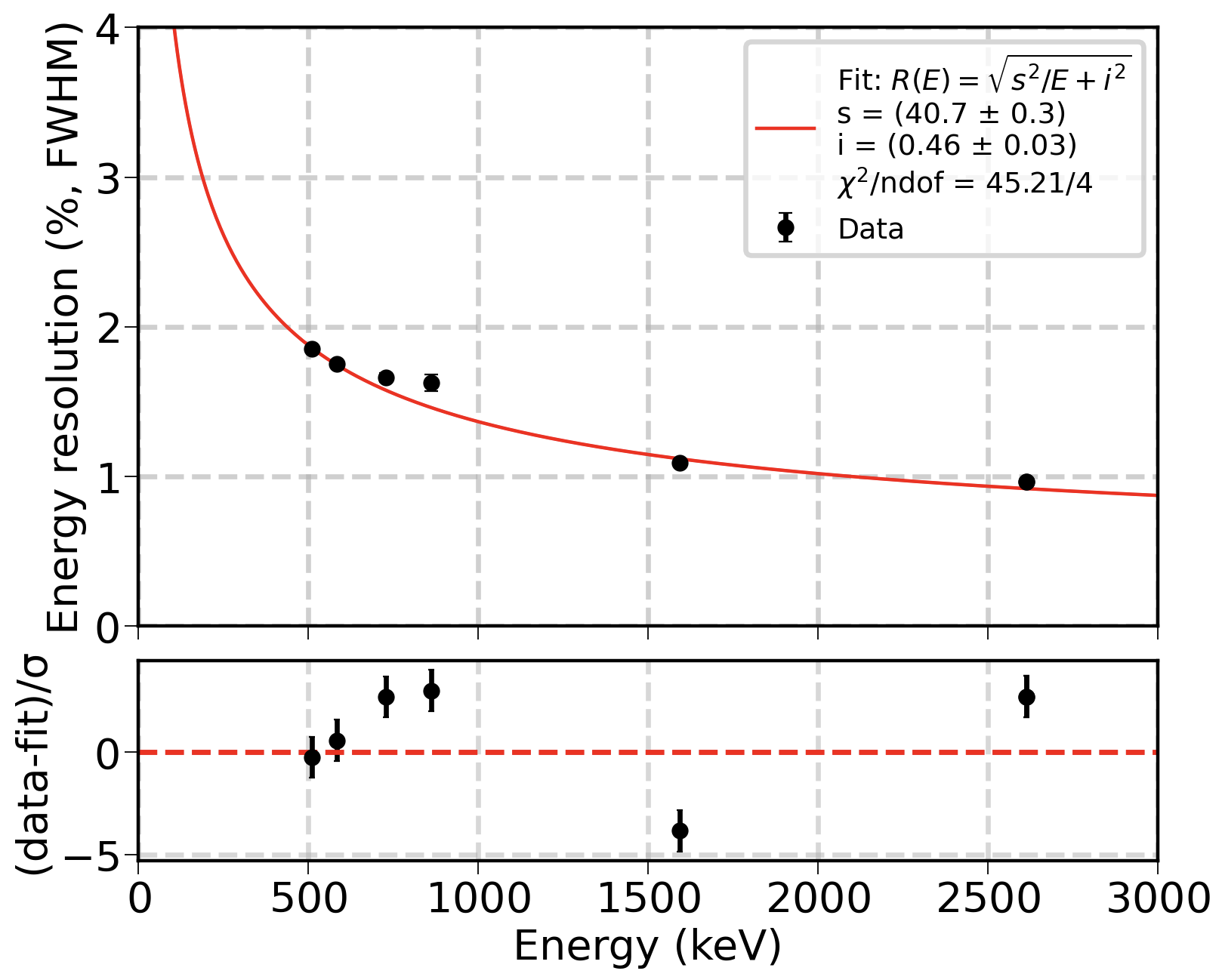}
    \caption{Plot of energy resolution $R$ FWHM in percentage as a function of the peak energy, shown together with the fitted curve and the corresponding residuals. In the legend, the fit results are also shown.}
    \label{fig:ereso}
\end{figure}

The energy resolution, expressed as $\mathrm{FWHM}/\mu$ in percent for each peak, is shown as a function of energy in Figure~\ref{fig:ereso}, together with the corresponding fit and residuals.

The data were fitted with the function
\[
R(E) = \sqrt{\frac{s^2}{E} + i^2},
\]
where the first term accounts for stochastic fluctuations that contributes to the energy resolution, while the second term represents the intrinsic limit of the detector resolution imposed by fundamental physical processes. In the equation, the term with $\sqrt{1/E^2}$ was neglected, since electronic noise is not expected to contribute significantly to the energy resolution in this energy range. This assumption is supported by a test fit in which the coefficient of the $\sqrt{1/E^2}$ term was found to be compatible with zero. The best-fit parameters are $s = (40.7 \pm 0.3)$~keV$^{1/2}\ \%$ and $i = (0.46 \pm 0.03)\ \%$. From the fitted curve, the energy resolution extrapolated to the \Qbb\ value is
\[
R(\Qbb) = (0.939 \pm 0.008)\%\ \mathrm{FWHM}.
\]

\section{Measurement of the 2615~keV $\gamma$ Photopeak}
\label{sec.photopeak}

Neutrinoless and two-neutrino double beta decay (\bbonu, \bbtnu) events involve the emission of two electrons from a single nucleus. In most cases, no additional tracks are produced (since no de-excitation X-ray is emitted in this case, only the eventual emission of Bremsstrahlung radiation produces extra energy deposits). Consequently, the resulting topology in the detector corresponds to a single continuous track. In contrast, $\mathrm{^{208}Tl}$ $\gamma$ rays tend to generate multiple clusters through Compton scattering and Bremsstrahlung emission, and photo-absorption of $\mathrm{^{214}Bi}$ $\gamma$ is accompanied in most cases by the emission of an X-ray. Therefore, to maximize the signal-to-background ratio in \bbonu\ and \bbtnu\ analyses, as shown in Ref.~\cite{PhysRevC.105.055501}, only single-track events are considered. 

\begin{figure}[t]
    \centering
    \includegraphics[width=.9\columnwidth]{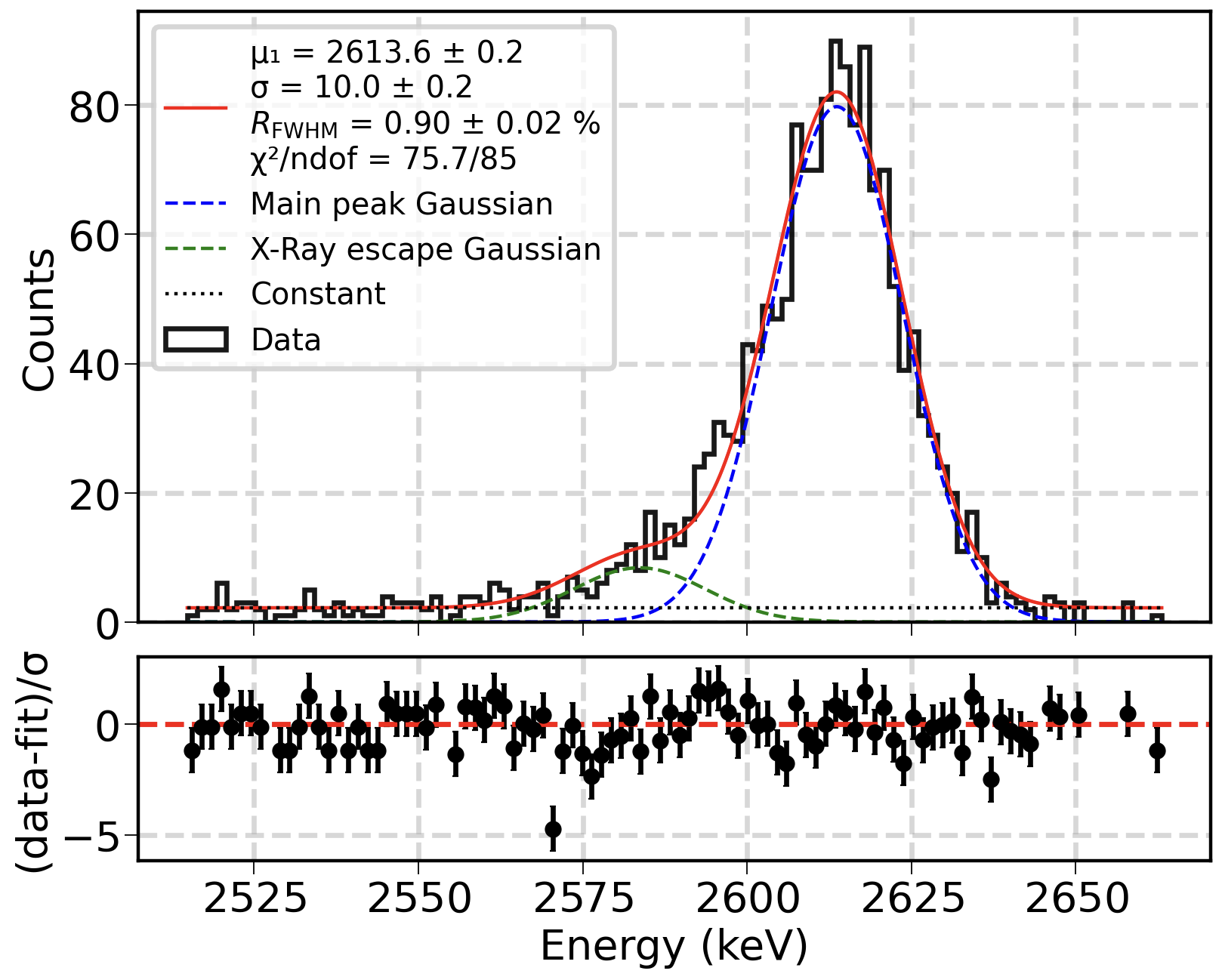}
    \caption{Single-cluster energy spectrum at the $^{208}$Tl photopeak. The fitted function with its different components, fit parameters, and residuals are also shown.}
    \label{fig:fitppsc}
\end{figure}

Accordingly, the energy resolution has also been evaluated using single-track events within the $\mathrm{^{208}Tl}$ photopeak. For this analysis, in addition to the selection criteria described in section~\ref{ss:ev_selection}, events are required to contain only one reconstructed cluster. The resulting energy spectrum for the $^{208}$Tl photoelectric peak is shown in Figure~\ref{fig:fitppsc}.

As seen in the figure, the escape-peak Gaussian component becomes more pronounced under the single-cluster selection, which preferentially retains events where a xenon de-excitation X-ray escapes the detector. The fit function used is the same as in Eq.~\ref{eq:fitppsc}, except that the error-function term is omitted, since multiple Compton events producing multiple clusters are excluded by this selection. 

With this requirement, an energy resolution of $R = (0.90 \pm 0.02)\%$ FWHM is obtained at the $^{208}$Tl photopeak. The energy resolution at \Qbb\ can be directly extrapolated from the $1/\sqrt{E}$ scaling (an excellent approximation given the proximity of both peaks) yielding $R(Q_{\beta\beta}) = (0.93 \pm 0.02)\%$ FWHM. 
\section{Measurement of the 2615~keV $\gamma$ Single-Escape Peak (SEP)}
\label{sec:sep}
\begin{figure*}[t]
    \centering
    \includegraphics[width=0.9\textwidth]{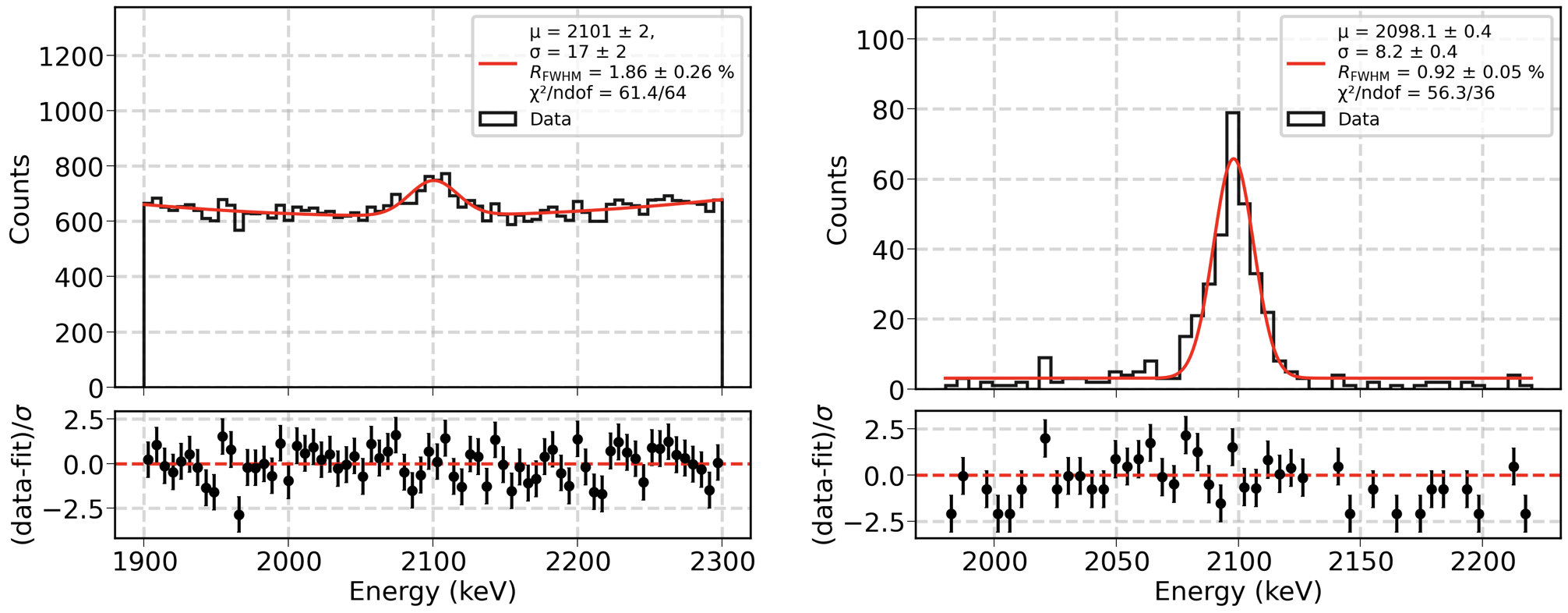}
    \caption{Left: Energy spectrum around the single-escape peak without any additional selection beyond the standard analysis cuts. Right: Same spectrum after requiring two clusters per event, with one of them having an energy compatible with 511 keV.}
    \label{fig:SEProw}
\end{figure*}

The mean free path of 511~keV $\gamma$ rays in xenon at 4~bar is approximately 5~m, much larger than the detector dimensions. Consequently, in the process
\[ 
\gamma(2615) + \mathrm{Xe} \;\to\; \mathrm{Xe} + e^- + e^+, 
\]
the two 511~keV $\gamma$ rays produced by positron annihilation are very likely to escape the detector. This makes the double-escape peak (DEP) a prominent feature, clearly visible in the \THO\ spectrum (see Figure \ref{fig:fits_le}). Conversely, the case in which only one of the two annihilation photons is contained while the other escapes is much less probable. As a result, the single-escape peak (SEP) is largely submerged in the Compton background and is barely visible, as shown in the left panel of Figure~\ref{fig:SEProw}. The spectrum was fitted using a second-degree polynomial to describe the Compton background and a Gaussian function to model the SEP signal.

However, SEP events exhibit a distinctive topological signature: a primary track accompanied by a spatially separated energy cluster of 511~keV. To identify such events, we require that two distinct clusters be reconstructed in the event, with one of them having an energy within $5\sigma$ of the mean value of the 511~keV peak:
\begin{equation}
E \in \left[\,511~\mathrm{keV} \pm 5\,\sigma(511~\mathrm{keV})\,\right].
\end{equation}
Applying this selection, the spectrum of the total reconstructed energy (sum of the two clusters) is shown in the right panel of Figure~\ref{fig:SEProw} together with the fit results and the plot of the residuals. For the background-suppressed data, the fit function used is the sum between a Gaussian and a flat term.
As illustrated in the Figure, this topological selection effectively suppresses the background and clearly reveals the SEP structure. The fit yields an energy resolution at the SEP of $R = (0.92 \pm 0.05)\%$ FWHM, fully consistent with the trend reported in figure \ref{fig:ereso}.
\section{Conclusions}\label{s:summ}
Relying on excellent energy resolution is crucial for \bbonu\ searches, as it directly determines the detector’s sensitivity. 

After installation at the Canfranc Underground Laboratory, NEXT-100 underwent low- and high-energy calibrations with \KR\ and \THO\ sources, respectively, to characterize its response across the full energy range and around the \Qbb\ value. 
Data were acquired under standard operating conditions, applying minimal selections to ensure clean S1–S2 association and full fiducial containment. 
Spatial and temporal variations in light yield were corrected using a 3D response map and simultaneous \KR\ calibration.

The detector response can be well approximated by a second degree polynomial with a very small quadratic correction from linearity and a discrepancy between measured and true energy below 0.1\% over the 511–2615 keV range.
Energy resolutions of $R = (1.09 \pm 0.01)\%$ FWHM at the \TL\ double escape peak and $R = (0.96 \pm 0.02)\%$ FWHM at the photopeak were obtained, yielding an extrapolated resolution from the fitted energy resolution curve at \Qbb\ of $R(Q_{\beta\beta}) = (0.939 \pm 0.008)\%$ FWHM—improving over the 1 \% design goal of the experiment.

When applying the single-cluster selection relevant for \bbonu\ and \bbtnu\ analyses, the energy resolution improves to $R = (0.90 \pm 0.02)\%$ FWHM at the $^{208}$Tl photopeak, corresponding to an extrapolated $R(Q_{\beta\beta}) = (0.93 \pm 0.02)\%$ FWHM. 
This energy resolution value confirms the exceptional performance of the NEXT technology for \bbonu\ searches.

\begin{acknowledgements}
The NEXT Collaboration acknowledges support from the following agencies and institutions: the European Research Council (ERC) under Grant Agreement No. 951281-BOLD and 101039048-GanESS; the European Union’s Framework Programme for Research and Innovation Horizon 2020 (2014–2020) under Grant Agreement No. 860881-HIDDeN; the MCIN/AEI of Spain and ERDF A way of making Europe under grants PID2021-125475NB and RTI2018-095979, and the Severo Ochoa and Mar\'ia de Maeztu Program grants CEX2023-001292-S, CEX2023-001318-M and CEX2018-000867-S; the Generalitat Valenciana of Spain under grants PROMETEO/2021/087, ASFAE/2022/028, ASFAE/2022/029, CISEJI/2023/27 and CIDEXG/2023/16; the Department of Education of the Basque Government of Spain under the predoctoral training program non-doctoral research personnel; the Spanish la Caixa Foundation (ID 100010434) under fellowship code LCF/BQ/PI22/11910019; the Portuguese FCT under project UID/FIS/04559/2020 to fund the activities of LIBPhys-UC; the Israel Science Foundation (ISF) under grant 1223/21; the Pazy Foundation (Israel) under grants 310/22, 315/19 and 465; the US Department of Energy under contracts number DE-AC02-06CH11357 (Argonne National Laboratory), DE-FG02-13ER42020 (Texas A\&M), DE-SC0019054 (Texas Arlington) and DE-SC0019223 (Texas Arlington); the US National Science Foundation under award number NSF CHE 2004111; the Robert A Welch Foundation under award number Y-2031-20200401. Finally, we are grateful to the Laboratorio Subterr\'aneo de Canfranc for hosting and supporting the NEXT experiment. C. Romo-Luque acknowledges financial support from LANL to participate in NEXT operations
\end{acknowledgements}

\bibliographystyle{spphys}
\bibliography{pool/NextRefs}

@article{NEXT:2012lrw,
    author = "Alvarez, V. and others",
    collaboration = "NEXT",
    title = "{Near-Intrinsic Energy Resolution for 30 to 662 keV Gamma Rays in a High Pressure Xenon Electroluminescent TPC}",
    eprint = "1211.4474",
    archivePrefix = "arXiv",
    primaryClass = "physics.ins-det",
    doi = "10.1016/j.nima.2012.12.123",
    journal = "Nucl. Instrum. Meth. A",
    volume = "708",
    pages = "101--114",
    year = "2013"
}

@ARTICLE{2013JInst,
       author = "Alvarez, V. and others",
        title = "{Initial results of NEXT-DEMO, a large-scale prototype of the NEXT-100 experiment}",
        collaboration = "NEXT",
      journal = {Journal of Instrumentation},
         year = 2013,
        month = apr,
       volume = {8},
       number = {4},
          eid = {P04002},
        pages = {P04002},
          doi = {10.1088/1748-0221/8/04/P04002},
archivePrefix = {arXiv},
       eprint = {1211.4838},
 primaryClass = {physics.ins-det},
       adsurl = {https://ui.adsabs.harvard.edu/abs/2013JInst...8P4002A}
}

@article{NEXT:2018rgj,
    author = "Monrabal, F. and others",
    collaboration = "NEXT",
    title = "{The Next White (NEW) Detector}",
    eprint = "1804.02409",
    archivePrefix = "arXiv",
    primaryClass = "physics.ins-det",
    reportNumber = "FERMILAB-PUB-18-113-CD",
    doi = "10.1088/1748-0221/13/12/P12010",
    journal = "JINST",
    volume = "13",
    number = "12",
    pages = "P12010",
    year = "2018"
}

@article{NEXT:2021vzd,
    author = "Sim{\'o}n, A. and others",
    collaboration = "NEXT",
    title = "{Boosting background suppression in the NEXT experiment through Richardson-Lucy deconvolution}",
    eprint = "2102.11931",
    archivePrefix = "arXiv",
    primaryClass = "physics.ins-det",
    reportNumber = "FERMILAB-PUB-21-114-SCD",
    doi = "10.1007/JHEP07(2021)146",
    journal = "JHEP",
    volume = "07",
    pages = "146",
    year = "2021"
}

@article{PhysRevC.105.055501,
  title = {Measurement of the $^{136}\mathrm{Xe}$ two-neutrino double-$\ensuremath{\beta}$-decay half-life via direct background subtraction in NEXT},
  author = {Novella, P. and others},
  collaboration = {NEXT},
  journal = {Phys. Rev. C},
  volume = {105},
  issue = {5},
  pages = {055501},
  numpages = {8},
  year = {2022},
  month = {May},
  publisher = {American Physical Society},
  doi = {10.1103/PhysRevC.105.055501},
  url = {https://link.aps.org/doi/10.1103/PhysRevC.105.055501}
}

@article{NEXT:2023daz,
    author = "Novella, P. and others",
    collaboration = "NEXT",
    title = "{Demonstration of neutrinoless double beta decay searches in gaseous xenon with NEXT}",
    eprint = "2305.09435",
    archivePrefix = "arXiv",
    primaryClass = "nucl-ex",
    reportNumber = "FERMILAB-PUB-23-251-ND",
    doi = "10.1007/JHEP09(2023)190",
    journal = "JHEP",
    volume = "09",
    pages = "190",
    year = "2023"
}

@article{NEXT:2025yqw,
    author = "Adams, C. and others",
    collaboration = "NEXT",
    title = "{The NEXT-100 Detector}",
    eprint = "2505.17848",
    archivePrefix = "arXiv",
    journal = {Approved at EPJC 10.1140/epjc/s10052-025-14951-y},
    month = "5",
    year = "2025"
}

@article{PhysRevLett.120.132501,
  title = {First Results from CUORE: A Search for Lepton Number Violation via $0\ensuremath{\nu}\ensuremath{\beta}\ensuremath{\beta}$ Decay of $^{130}\mathrm{Te}$},
  author = {Alduino, C. and others},
  collaboration = {CUORE},
  journal = {Phys. Rev. Lett.},
  volume = {120},
  issue = {13},
  pages = {132501},
  numpages = {8},
  year = {2018},
  month = {Mar},
  publisher = {American Physical Society},
  doi = {10.1103/PhysRevLett.120.132501},
  url = {https://link.aps.org/doi/10.1103/PhysRevLett.120.132501}
}

@article{Agostini_2021,
   title={Calibration of the Gerda experiment},
   collaboration = {GERDA},
   volume={81},
   ISSN={1434-6052},
   url={http://dx.doi.org/10.1140/epjc/s10052-021-09403-2},
   DOI={10.1140/epjc/s10052-021-09403-2},
   number={8},
   journal={The European Physical Journal C},
   publisher={Springer Science and Business Media LLC},
   author={Agostini, M.  and others},
   year={2021},
   month=aug }

@article{Mart_nez_Lema_2018,
   title={Calibration of the NEXT-White detector using 83mKr decays},
   volume={13},
   collaboration = {NEXT},    
   ISSN={1748-0221},
   url={http://dx.doi.org/10.1088/1748-0221/13/10/P10014},
   DOI={10.1088/1748-0221/13/10/p10014},
   number={10},
   journal={Journal of Instrumentation},
   publisher={IOP Publishing},
   author={Martínez-Lema, G. and others},
   year={2018},
   month=oct, pages={P10014–P10014} }

@article{Resowhite,
   title={Energy calibration of the NEXT-White detector with 1\% resolution near $\ensuremath{Q}_{\ensuremath{\beta}\ensuremath{\beta}}$ of $^{136}\mathrm{Xe}$},
   collaboration = {NEXT}, 
   volume={2019},
   ISSN={1029-8479},
   url={http://dx.doi.org/10.1007/JHEP10(2019)230},
   DOI={10.1007/jhep10(2019)230},
   number={10},
   journal={Journal of High Energy Physics},
   publisher={Springer Science and Business Media LLC},
   author={Renner, J. and others},
   year={2019},
   month=oct }

@article{Renner_2018,
   title={Initial results on energy resolution of the NEXT-White detector},
   collaboration = {NEXT},
   volume={13},
   ISSN={1748-0221},
   url={http://dx.doi.org/10.1088/1748-0221/13/10/P10020},
   DOI={10.1088/1748-0221/13/10/p10020},
   number={10},
   journal={Journal of Instrumentation},
   publisher={IOP Publishing},
   author={Renner, J. and others},
   year={2018},
   month=oct, pages={P10020–P10020} }

@article{DackRejNN,
   title={Demonstration of background rejection using deep convolutional neural networks in the NEXT experiment},
   volume={2021},
   collaboration = {NEXT},
   ISSN={1029-8479},
   url={http://dx.doi.org/10.1007/JHEP01(2021)189},
   DOI={10.1007/jhep01(2021)189},
   number={1},
   journal={Journal of High Energy Physics},
   publisher={Springer Science and Business Media LLC},
   author={Kekic, M. and others},
   year={2021},
   month=jan }

@article{EvIDNEXTW,
   title={Demonstration of the event identification capabilities of the NEXT-White detector},
   collaboration = {NEXT},
   volume={2019},
   ISSN={1029-8479},
   url={http://dx.doi.org/10.1007/JHEP10(2019)052},
   DOI={10.1007/jhep10(2019)052},
   number={10},
   journal={Journal of High Energy Physics},
   publisher={Springer Science and Business Media LLC},
   author={Ferrario, P. and others},
   year={2019},
   month=oct }

@article{BkgNW1,
    author = "Novella, P. and others",
    collaboration = "NEXT",
    title = "{Measurement of radon-induced backgrounds in the NEXT double beta decay experiment}",
    eprint = "1804.00471",
    archivePrefix = "arXiv",
    primaryClass = "physics.ins-det",
    reportNumber = "FERMILAB-PUB-18-093-CD-ND",
    doi = "10.1007/JHEP10(2018)112",
    journal = "JHEP",
    volume = "10",
    pages = "112",
    year = "2018"
}

@article{NEXT:2025fpq,
  author        = {Martínez-Lema, G. and others},
  collaboration = {NEXT},
  title         = {First results of the NEXT-100 detector using {${}^{83\mathrm{m}}$Kr} decays},
  journal       = {arXiv preprint},
  note          = {arXiv:2511.01710 [hep-ex]},
eprint = "2511.01710",
archivePrefix = "arXiv",
primaryClass = "hep-ex",
reportNumber = "FERMILAB-PUB-25-0878",
month = "11",
year = "2025"
}

@article{BkgNW2,
   title={Radiogenic backgrounds in the NEXT double beta decay experiment},
   collaboration = "NEXT",
   volume={2019},
   ISSN={1029-8479},
   url={http://dx.doi.org/10.1007/JHEP10(2019)051},
   DOI={10.1007/jhep10(2019)051},
   number={10},
   journal={Journal of High Energy Physics},
   publisher={Springer Science and Business Media LLC},
   author={Novella, P. and others},
   year={2019},
   month=oct }

\end{document}